\documentclass{article}
\usepackage[utf8]{inputenc}
\usepackage{graphicx}
\usepackage{authblk}
\usepackage{cite}

\title{Solid-State Biased Coherent Detection of Ultra-Broadband Terahertz Pulses for high repetition rate, low pulse energy lasers}
\date{} 

\author[1]{Tim Suter}
\author[2]{Alessandro Tomasino}
\author[1]{Matteo Savoini}
\author[3]{Sarah Houver}
\author[1,4]{Steven L. Johnson}
\author[1]{Elsa Abreu}

\affil[1]{Institute for Quantum Electronics, Department of Physics, ETH Zürich, 8093 Zürich, Switzerland}
\affil[2]{INRS-EMT, 1650 Boulevard Lionel-Boulet, Varennes, QC J3X 1S2, Canada}
\affil[3]{Laboratoire Matérieux et Phénomènes Quantiques, Université de Paris, CNRS, 75013 Paris, France}
\affil[4]{Laboratory for Non Linear Optics, Paul Scherrer Institute, 5232 Villigen PSI, Switzerland}

\begin{document}

\maketitle

\begin{abstract}
We report the coherent generation and detection of terahertz (THz) pulses covering the bandwidth of \mbox{0.1-9 THz} in a high repetition rate, low pulse energy laser system.
In this work we demonstrate the application and evaluation of solid-state biased coherent detection in combination with a spintronic emitter. This combination was used to generate and detect THz pulses in a time-domain spectroscopy (TDS) setup and tested on bulk nonlinear crystals. These results establish a new promising candidate to extend the possibilities for compact, broadband THz TDS systems driven by high repetition rate lasers.
\end{abstract}

\section{Introduction}
Terahertz (THz) radiation is situated between microwave and infrared radiation and has historically been difficult to generate and detect, resulting in what is often called the THz gap. The advent of modelocked lasers systems, in particular Ti:Sapphire lasers, has opened up the possibility to coherently generate and detect THz radiation by using ultrashort pulses. Time-Domain Spectroscopy (TDS) provides access to the detection of both amplitude and phase of THz pulses, which enables for example contact-less measurement of conductivity, excitation dynamics and phase transitions in materials. Various applications of THz TDS have been established most notably in physics, material science, chemistry and biology \cite{lee-2008, jepsen-2005}. 
\\~\\
Standard detection methods for THz pulses include electro-optic (EO) sampling in nonlinear crystals, photoconductive antennas and air biased coherent detection (ABCD) \cite{lee-2008, jepsen-2005}. Both EO sampling and photoconductive anntennas are suitable to investigate the lower THz frequencies and their bandwidth is mainly limited by phonon absorption \cite{burford-2017, wu-1996, leitenstorfer-1999}. In the case of EO sampling, commonly used nonlinear crystals ZnTe and GaP are limited to \mbox{3.5 THz} and \mbox{8 THz} respectively. This limitation can partially be avoided by resorting to very thin crystals, which however lead to an early Fabry-Pérot reflection of the THz pulses in the crystals. As the detection sensitivity scales linearly with crystal thickness, very thin crystals require large THz fields, limiting their use in high repetition rate setups.\\
The ABCD technique relies on the electric field induced second harmonic generatrion (EFISH) in gases and is therefore not affected by phonon absorption. The EFISH process is a third order nonlinear interaction between an optical probe pulse, the THz field and an external bias field, resulting in a beam oscillating at the second harmonic of the probe beam. The experimental drawback of the ABCD technique is that it requires bias voltages on the order of kV and probe energies on the order of tens of µJ \cite{dai-2004,karpowicz-2008}. A new detection technique has recently been discovered, building upon the EFISH process used in ABCD. In solid-state biased coerent detection (SSBCD) the EFISH process takes place in a thin layer of silicon nitride (SiN) instead of a gas \cite{tomasino-2017, tomasino-2018, tomasino-2021}. The benefit is a much larger $\chi^{(3)}$ coefficient compared to the gases used in ABCD, which allows the use of bias voltages and probe energies several orders of magnitude lower compared to ABCD. Additionally, it can be manufactured into a single compact device, whereas for ABCD a gas cell is necessary. The bias field in the SSBCD device is generated in a 1 µm wide metallic slit. This slit size which is subwavelength for THz is beneficial as it leads to a THz field enhancement in the slit of up to a factor of 6.
\\~\\
In the generation of coherent THz pulses similar bandwidth issues arise as in the detection. Standard options are optical rectification, photoconductive antennas and two-color plasma sources. Optical rectification in nonlinear crystals and photoconductive antennas suffer from the same phonon absorptions that limit the detection \cite{lee-2008, jepsen-2005}. Plasma sources generate THz pulses which can cover and even go beyond the whole \mbox{0.1-10 THz} gap, but they require ultrashort pulses with pulse energies on the order of a few mJ \cite{cook-2000, thomson-2010, clerici-2013}. In recent years, a novel approach has emerged in the form of spintronic emitters capable of covering the range from \mbox{0.1-30 THz} \cite{seifert-2016, seifert-2017}. In such an emitter, ultrashort pulses lead to demagnetization of a ferromagnetic layer, which in turn generates a spin-current into a nonferromagnetic layer. Through the inverse spin-Hall effect the spin-currents are converted into charge currents, leading to the emission of a THz pulse. The performance of the spintronic emitter is further improved by employing a trilayer structure where a nonmagnetic layer sits between two ferromagnetic layers. 
\\~\\
In high field setups, the combination of two-color plasma sources with ABCD has been established to cover the \mbox{0.1-10 THz} range and can be further simplified through the use of SSBCD \cite{tomasino-2017, tomasino-2018, tomasino-2021}. In the context of low field setups, we show here that the combination of a spintronic emitter and SSBCD is a very promising approach to obtain a system covering the entire \mbox{0.1 - 10 THz} region. While the SSBCD technique has very successfully been used to replace an existing ABCD setup in high field setups driven by a plasma source at a repetition rate of 1 kHz \cite{tomasino-2017, tomasino-2018, tomasino-2021}, its suitability for setups with high repetition rates and low THz fields has not yet been demonstrated. In this work we examine the performance of a SSBCD device in a low THz field setup with a repetition rate of \mbox{250 kHz} and use it to study three test samples. We also discuss the installation and alignment procedure, which are significantly more challenging than when the SSBCD device is simply placed at the position of a working ABCD setup.
\section{Experimental methods}
A \mbox{40 fs}, \mbox{800 nm}, \mbox{250 kHz}, \mbox{6.2 µJ} pulsed laser was used to drive a THz TDS setup. The THz radiation was generated using a trilayer spintronic emitter (\mbox{2 nm W}, \mbox{1.8 nm CoFeB}, \mbox{2 nm Pt} on \mbox{500 nm} of Sapphire). Such an emitter has been shown to generate \mbox{~30 THz} wide pulses measured by EO sampling in a \mbox{70 µm} thick LAPC crystal, although with shorter pump pulse duration \cite{seifert-2016}. A thin layer of PTFE was used to block the 800 nm residue. A peak THz amplitude of \mbox{4.5 kV/cm} was measured with EO sampling in a \mbox{300 µm GaP} crystal in a nitrogen-purged environment. 
\\~\\
The SSBCD device was purchased from Ki3 Photonics \cite{ki3photonics}. It consists of a 500 µm thick silica substrate, a 1 µm wide gold slit filled with SiN and a 1 µm thick SiN cover layer. The design is described in great detail in \cite{tomasino-2017, tomasino-2018}. 
The interaction of the THz, bias and probe field in the SiN layer generates radiation oscillating at the second harmonic of the probe field according to the EFISH process \cite{tomasino-2017}. The intensity of the generated beam is given by
\begin{equation}
I_{SH}^{total} \propto (\chi^{(3)} I_\omega)^2[(E_{THz})^2 + (E_{bias})^2  \pm 2E_{THz}E_{bias}],
\label{EFISH}
\end{equation}
where $\chi^{(3)}$ is the third order susceptibility of the nonlinear medium, $I_\omega$ is the probe beam intensity, $E_{THz}$ and $E_{bias}$ denote the electric field of the THz pulse and bias field respectively. The double sign is determined by the polarity of the bias field and the relative phase between the bias field and the THz field. By subtracting the intensities acquired from both polarities, we get a value proportional to the THz field, therefore we can reconstruct the amplitude and the phase of the THz field.
Using a function generator and a high voltage amplifier we applied rectangular AC voltages with a frequency of 500 Hz. A photomultiplier tube was used to measure the second harmonic beam generated by the EFISH process and a preamplifier was used for noise reduction. The function generator was also used to trigger the data acquisition in a DAQ card, which allowed us to determine the bias polarity with which a measurement was taken. This is important, as it is then not required to synchronize the AC bias switching to the 250 kHz laser repetition rate, which is much larger than the frequency of the AC bias switching. An offset in second harmonic intensity between the two voltage polarities was observed even in the absence of a THz field. To reduce noise originating from sudden changes in voltages, which are believed to be caused by discharge effects, the time traces from both polarities were then scaled to the average intensity without a THz field. To evaluate the statistical reliability of our measurements we determined the dynamic range (DR) and signal to noise ratio (SNR). We defined the SNR to be the ratio of the mean peak value to the standard deviation of the peak value and the DR to be the ratio of the mean peak value to the root mean square of the noise floor \cite{naftaly-2009}.
\\~\\
The alignment critically depends on spatial overlap between the focal points of the probe and THz beam into the \mbox{1 µm} wide slit of the device. Achieving spatial overlap only with the SSBCD device is extremely tedious and time-consuming, since a small misalignment can already lead to a complete disappearance of the THz signal from the SSBCD device. We have therefore developed a procedure such that we could make use of pre-alignment of the system using EO sampling. In the SSBCD detection configuration, THz and probe beam are focused by a \mbox{2 inch} parabolic mirror and a \mbox{125 mm} focal length lens, respectively. This short focal length for the probe beam is, however, not compatible with EO sampling so that the beams were first overlapped using a long focal length lens. A beam profiling camera placed at the EO sampling crystal position was then used to preserve the alignment and focal position of the probe beam upon switching back to the short focal length lens. Following this method, we already start with an alignment where only very small optimization of the spatial overlap on the SSBCD device is necessary.
\section{Results}
\begin{figure}[htbp]
\centering\includegraphics[width=9cm]{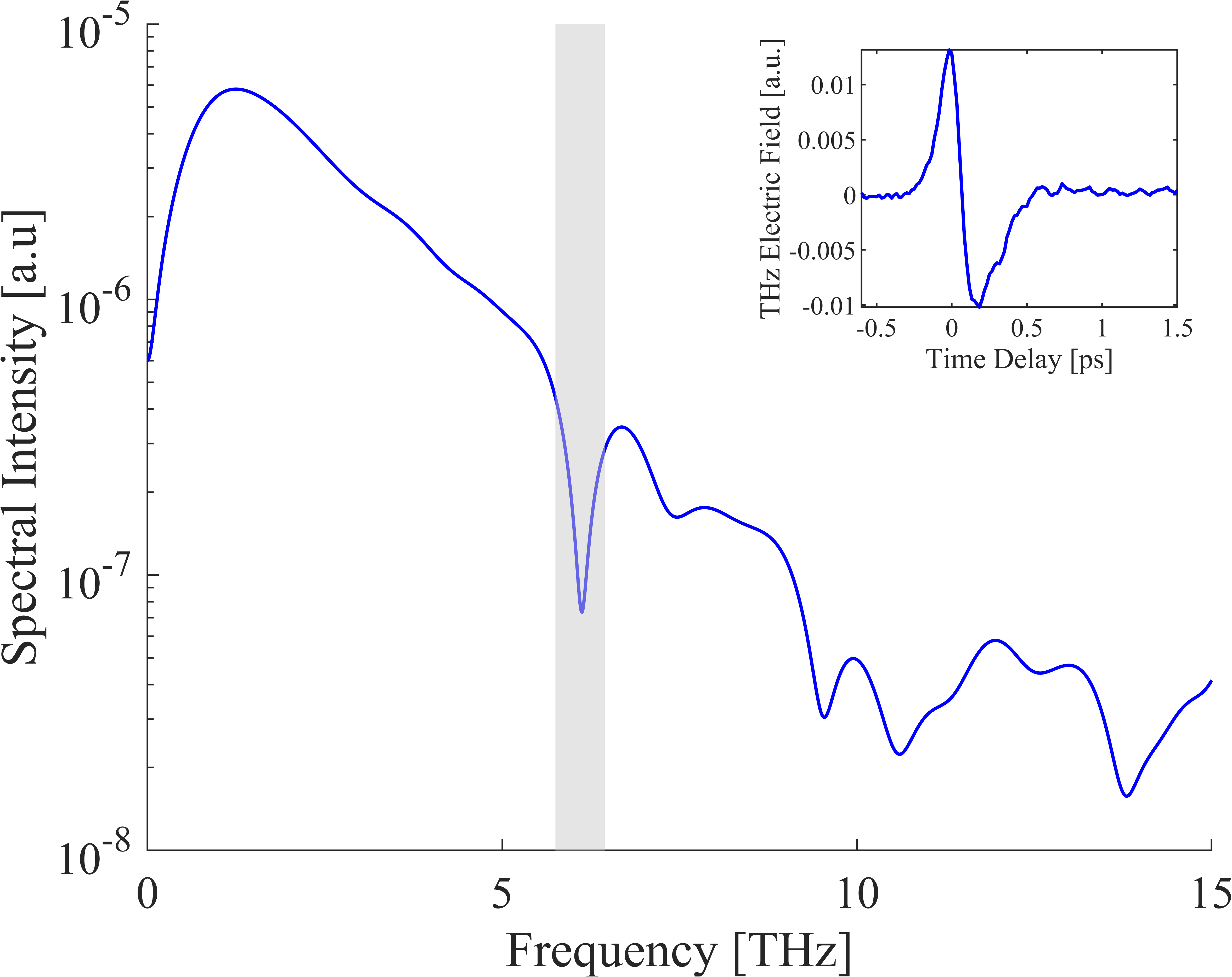}
\caption{THz spectrum acquired with \mbox{32 nJ} probe pulses, \mbox{$\pm$30 V} bias voltage. The corresponding time-trace is shown in the inset. The greyed out region around \mbox{6.1 THz} indicates the PTFE absorption.}
\label{Figure 1}
\end{figure}

A spectrum and the corresponding time trace measured with the SSBCD device are shown in Figure \ref{Figure 1}. A time-window was used to smoothly bring the edges of the trace to zero before performing the Fourier transform. The time trace was obtained with \mbox{32 nJ} probe energy and \mbox{30 V} bias voltage. 
To confirm the field dependencies stated in the EFISH equation (\ref{EFISH}) we measured traces using different probe energies and bias voltages. The probe energy was varied from \mbox{8 nJ} to \mbox{32 nJ} while the bias voltage was varied from \mbox{25 V} to \mbox{42 V}. The error bars correspond to the standard deviation of the peak THz signal for an average of 10 measurements and integration time of \mbox{0.5 s}. A quadratic fit was applied to the probe energy dependence and a linear fit was applied to the bias voltage dependence.
\begin{figure}[htbp]
\centering\includegraphics[width=10cm]{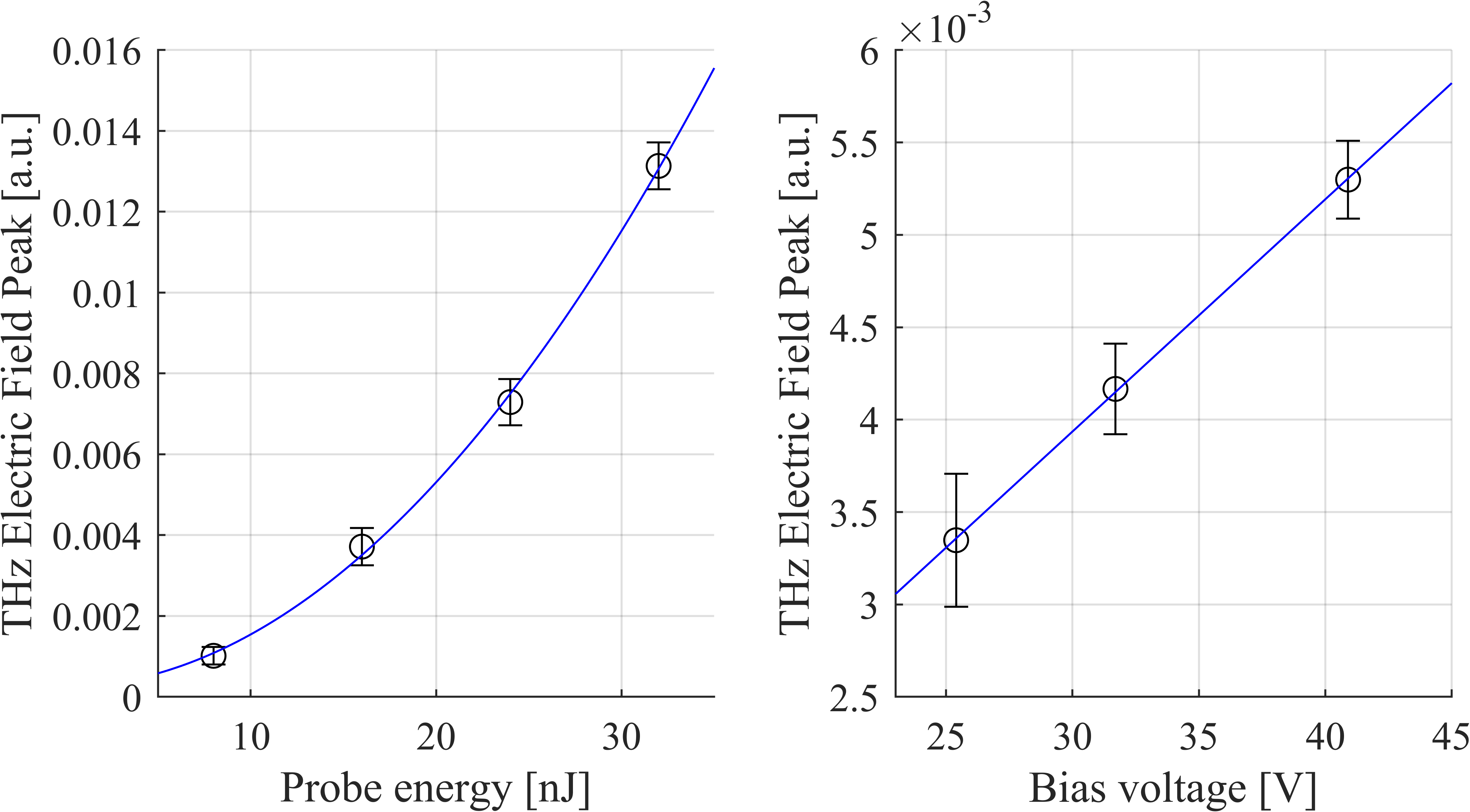}
\caption{Comparison of measured values of the THz electric field for a) different probe energies with constant \mbox{30 V} bias voltage and b) different bias voltages with constant \mbox{16 nJ} probe energy.}
\label{Figure 2}
\end{figure}
\\
To test the performance of the system we investigated the transmission of various crystals with features in the \mbox{0.1-10 THz} range \cite{wu-2018,casalbuoni-2008}. The crystals were placed between THz emission and detection. The investigated crystals are ZnTe (\mbox{500 µm} thick), GaAs (\mbox{625 µm} thick) and GaP (\mbox{100 µm and 300 µm thick}) for which the measured transmission spectra are shown in Figure \ref{Figure 3}. The measurements are averaged from 20 measurements with an integration time of \mbox{0.5 s}. As a guide to the accuracy of the measurements shown in Figure \ref{Figure 3} we added the noise floor line, which was obtained by dividing the average noise floor value of the reference measurement by the reference spectrum.
\\~\\
A model of the dielectric function including only the response from the lowest TO phonon was used as follows:
\begin{equation}
\epsilon(f) = \epsilon_\infty + \frac{S_0 f^2}{f_0^2 - f^2 -i\Gamma_0f}
\label{TOPhonon}
\end{equation}
where we introduce the optical dielectric constant at high frequencies $\epsilon_\infty$ and $S_0$, $f_0$ and $\Gamma_0$ as the oscillator strength, eigenfrequency and damping constant, respectively. The corresponding material parameters \cite{wu-2018} are listed in table \ref{material coefficients}. 
\begin{table}[htbp]
    \centering
\begin{tabular}{|l|l|l|l|l|}
\hline
Crystal & $\epsilon_{\infty}$ & $S_0$ & $f_0$ (THz) & $\Gamma_0$ \\ \hline
     ZnTe     &    7.4       &      2.7     &       5.3    &    0.09       \\ \hline
        GaAs  &       11.1    &      1.8     &      8.025     &     0.066      \\ \hline
        GaP  &      8.7     &       1.8    &    10.98       &       0.02    \\ \hline
\end{tabular}
    \caption{Material parameters for ZnTe, GaAs and GaP.}
    \label{material coefficients}
\end{table}
\\
The calculated transmission following equation \ref{TOPhonon} is also shown in Figure \ref{Figure 3} for the four samples, using dashed lines.
\begin{figure}[htbp]
\centering\includegraphics[width=\linewidth]{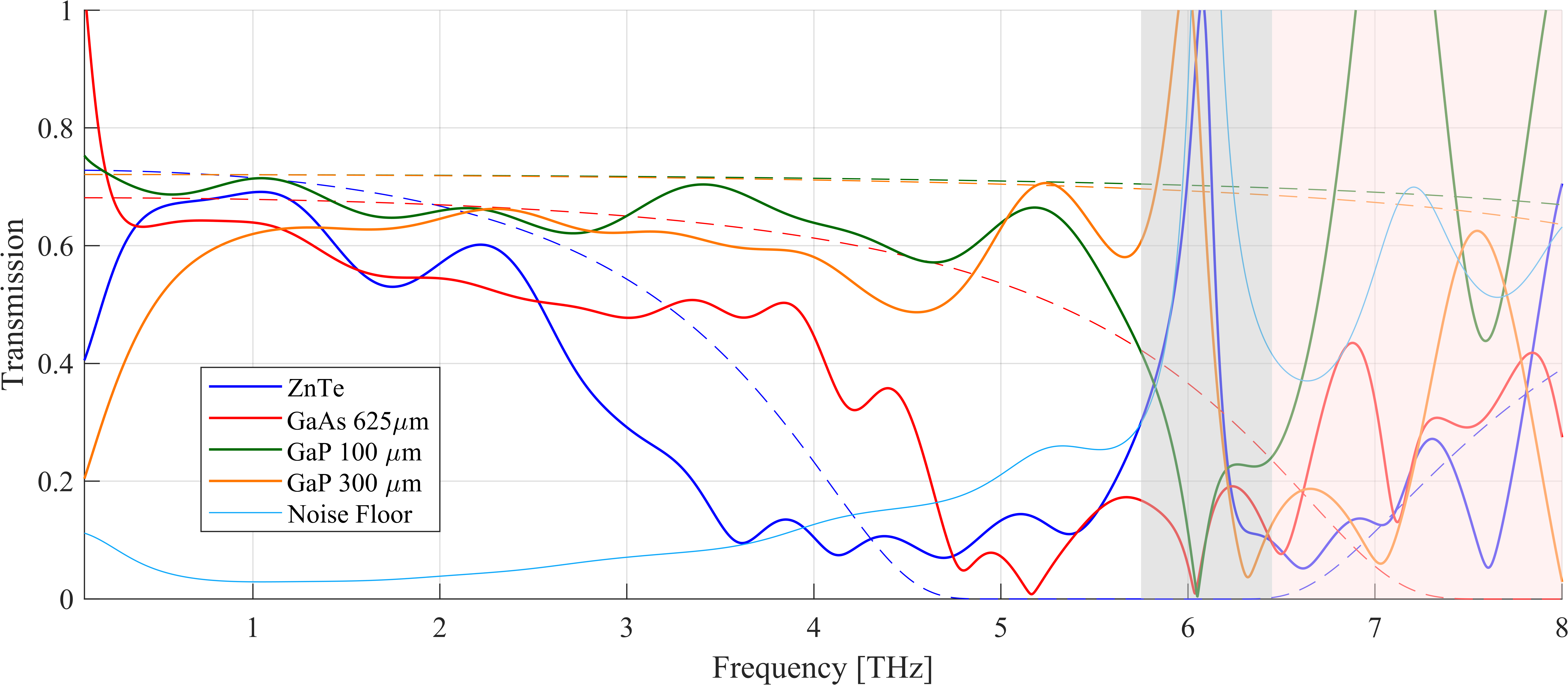}
\caption{Transmission of ZnTe \mbox{(500 µm)}, GaAs \mbox{(625 µm)} and GaP \mbox{(100 µm, 300 µm)}. The dashed lines show the transmission following the TO-Phonon model not including multiphonon processes. The area shaded red indicates where the transmission is too close to the noise floor for a reliable statement about the transmission.}
\label{Figure 3}
\end{figure}

\section{Discussion}
The spectrum shown in Figure \ref{Figure 1} shows the frequency components expected from the spintronic emitter used with pump pulse durations of \mbox{40 fs} up to at least \mbox{9 THz}. The sharp decrease around \mbox{6.1 THz} can be attributed to absorption in PTFE due to a CF2 twisting mode \cite{dangelo-2014}. This could be avoided by using a high resistivity silicon wafer instead of PTFE to separate the \mbox{800 nm} residue from the THz beam \cite{dai-2004}. The advantage of using PTFE is the very high transmission \mbox{(85-90\%)} below \mbox{6 THz}, contrary to Silicon which transmits only 70\% of the THz field \cite{dangelo-2014}. From the time trace we calculated a SNR of 23 and a DR of 66. While the SSBCD device is capable of a DR two orders of magnitudes larger, it is conventionally used with much larger probe energies (\mbox{50-100 nJ} compared to \mbox{32 nJ}) \cite{tomasino-2017, tomasino-2018, tomasino-2021}. As the DR and SNR scale quadratically with the probe energy, such low SNR and DR values are expected for the combination of low probe energy and low THz field used in this setup. \\
Not only have we shown that the SSBCD device can be used with low energy, high repetition rate lasers but we have also demonstrated that it can be setup using low THz fields only. While it has been shown before that the the SSBCD technique can be used to detect THz fields at least as low as \mbox{6 kV/cm} in a high field setup, this has been done by aligning first with a high field and then decreasing the field through a pair of wire grid polarizers.
\\~\\
The dependence of the THz peak field values in the time domain on both probe energy and bias voltage as shown in Figure \ref{Figure 2} agrees very well with the expected behaviour (quadratic and linear respectively). A significant decrease in THz peak field value over time has been observed for probe energies as low as \mbox{24 nJ}, indicating the onset of damage of the slit. This limits the sensitivity of our SSBCD setup, which would benefit greatly from larger probe energies. The damage threshold of the bias voltage was not investigated, as it could lead to permanent device failure. Compared to previous publications on the SSBCD technique \cite{tomasino-2017, tomasino-2018, tomasino-2021} we use significantly shorter pulse durations (\mbox{40 fs} compared to \mbox{155 fs}). While we expect the shorter pulse duration to be beneficial to the \mbox{SSBCD} technique, the larger peak intensities might contribute to the lower damage threshold in probe pulse energy. Another likely factor is that the high repetition rate does not leave enough time for the device to dissipate heat deposited by the laser. The average power of the probe beam is \mbox{8 mW} for \mbox{32 nJ} pulses at \mbox{250 kHz}, which is significantly larger than the average power of \mbox{0.1 mW} for \mbox{100 nJ} pulses at \mbox{1 kHz} used in previous publications \cite{tomasino-2017, tomasino-2018, tomasino-2021}. Due to the very small slit size, any thermal expansion effects might significantly impact the performance of the device.
\\~\\
The sample transmission shown in Figure \ref{Figure 3} is lower than the expected transmission based on the lowest TO phonon only. However the transmission through all of the samples studied in this work is additionally affected by multiphonon processes, which significantly contribute to absorption losses already at frequencies below the TO-phonon. For ZnTe there are distinctive two-phonon absorption processes at \mbox{1.6 THz} and \mbox{3.7 THz} \cite{lee-2008} which are present in Figure \ref{Figure 3}. The lowest TO-Phonon of GaP is at \mbox{11 THz}, so that the lower THz region should be insensitive to the sample thickness and mainly be affected by reflecion losses at the sample surfaces. This is, however, not the case due to the multiphonon processes which are known to occur at \mbox{3 THz} and \mbox{4.3 THz} respectively \cite{saito-2008}, which lead to a thickness dependent increase in absorption at those frequencies in the GaP spectra of Figure \ref{Figure 3}. For GaAs there are multiphonon processes at \mbox{2.4 THz} and \mbox{5 THz} \cite{stolen-1969} which start to significantly increase the absorption in addition to the TO-phonon mode absorption.  
The region around \mbox{6.1 THz} deviates from theory for all samples as the PTFE completely absorbs any THz radiation for both the sample and the reference measurements. The noise floor serves as a useful guide to the sensitivity of the measurement, especially at larger frequencies. Above the PTFE absorption frequency, the transmission for all samples has decreased to a level comparable to the noise floor, which prevents the determination of any true THz transmission above \mbox{6 THz}.
It is, however, worth noting that the spectrum of Figure \ref{Figure 1} extends up to \mbox{9 THz}. This indicates the presence of THz radiation and the ability to detect it using the SSBCD technique in the \mbox{0.1-9 THz} range. A small decrease in THz due to absorption or reflection loss at a sample will however quickly lead to THz fields that are too weak in the high frequency range to be detected by the setup in its current state. While this is not a constraint of the SSBCD technique in general it can be attributed to the low SNR and DR values due to the low probe energy.
\\~\\
Recently an alternative approach to the bias field was presented by applying a DC voltage instead of an AC voltage \cite{tomasino-2021}. By measuring two traces with opposite polarity one can reconstruct the THz trace. This approach was tried as well, however we could only partially reproduce the results from \cite{tomasino-2021}. A significant increase in second harmonic generation was observed when the bias voltage was applied, however this did not stabilize but instead decayed rapidly. It is as of yet unclear if this is a consequence of the higher repetition rate of the probe beam or if it is specific to our device and requires further investigation.
\\~\\
To improve upon our measurements we suggest further investigation into the design of the SSBCD device, specifically adapting it for high repetition rate lasers and shorter pulse durations to achieve a larger damage threshold. One possibility would be to investigate the effect of a larger slit size. This would lead to a smaller field enhancement in the slit. But this could be rapidly compensated by increasing the probe energy, since SNR and DR are expected to scale quadratically with probe energy \cite{tomasino-2018}. Another suggestion would be to investigate other nonlinear materials which might be more robust than SiN to the high repetition pulse exposure. Further setup optimization including stronger focusing of the THz beam might lead to slight performance improvements. A larger bias voltage would also improve the performance, which would require investigating the bias voltage damage threshold of the device under high repetition rate probe beam exposure, which we have not attempted for fear of inducing irreversible damage.
\\~\\
If the device could be adapted to achieve a larger damage threshold, then the sensitivity could be drastically increased. The combination of SSBCD with the broadband THz emission of the spintronic emitter is a very promising route to obtain a broadband THz TDS setup for low pulse energy, high repetition rate laser systems.
\section{Conclusion}
We have demonstrated that the the SSBCD technique can be used for coherent THz detection in combination with a high repetition rate, low energy laser setup. Through careful alignment it can be used without the need of high field THz pulses or an already existing ABCD setup. We have successfully shown its use in analyzing bulk samples of three different materials. Its main limitation is its low damage threshold, attributed to the high repetition rate and the short pulse duration. If this limitation could be overcome through a change in design or in the materials used in the SSBCD device, this technique could become a unique way to detect the \mbox{0.1-10 THz} range in THz TDS setups based on high repetition rate laser systems. The combination of SSBCD detection with THz generation in a spintronic emitter would then be drastically more broadband compared to common setups in which THz generation and detection are based on photoconductive antennas or ZnTe crystals.
\\~\\
\small
\textbf{Funding.}\quad
EA acknowledges funding from the Swiss National Science Foundation through Ambizione Grant PZ00P2 179691.\\[5pt]
\textbf{Acknowledgments.} \quad
The authors would like to thank Prof. T. Kampfrath (FU Berlin) for supplying us with the spintronic emitter. We also thank Prof. R. Morandotti (INRS-EMT), as well as Ki3 Photonics and Y. Jestin for their help and for providing us with the state-of-the-art SSBCD detection system.\\[5pt]
\textbf{Disclosures.} \quad
The authors declare no conflicts of interest.\\[5pt]
\textbf{Data availability.} \quad Data underlying the results presented in this paper will be made available in the ETH Research Collection upon publication.\\[5pt]
\textbf{Supplemental document.} \quad
A supplemental document containing detailed information on the individual bias voltages, traces and spectra of the data shown in \mbox{Figure 2} will be provided upon publication.\\~\\

\bibliography{main}{}
\bibliographystyle{unsrt}

\end{document}